\documentclass[a4paper,fleqn]{cas-dc}

\usepackage[numbers]{natbib}
\usepackage{multirow}
\usepackage{multicol}
\usepackage{csquotes}
\usepackage{amsmath}
\usepackage{amssymb}
\usepackage{algpseudocode}
\usepackage{graphicx}
\usepackage{subcaption}
\usepackage{footnote}
\usepackage{soul}
\usepackage{hyperref}
\usepackage{import}

\def\tsc#1{\csdef{#1}{\textsc{\lowercase{#1}}\xspace}}
\tsc{WGM}
\tsc{QE}
\tsc{EP}
\tsc{PMS}
\tsc{BEC}
\tsc{DE}

\begin{document}
\let\WriteBookmarks\relax
\def\floatpagepagefraction{1}
\def\textpagefraction{.001}

\shorttitle{The role of large language models in UI/UX design}

\title [mode = title]{The role of large language models in UI/UX design: A systematic literature review}                      

\shortauthors{A, Ahmed. et~al.}

\author[1]{Ammar Ahmed}[orcid=0009-0003-9984-4819, type=editor]

\ead{ammaa@stud.ntnu.no}

\credit{Conceptualization, methodology, data collection, original draft preparation, writing}

\author[1]{Ali Shariq Imran}[orcid=0000-0002-2416-2878, type=editor]
\ead{ali.imran@ntnu.no}
\cormark[1]

\credit{Methodology, review and editing, supervision}


\affiliation[1]{organization={Intelligent Systems and Analytics (ISA) Research Group, Department of Computer Science (IDI), Norwegian University of Science \& Technology (NTNU)},
    city={Gj\o vik},
    postcode={2815}, 
    country={Norway}}


\begin{abstract}
  This systematic literature review examines the role of large language models (LLMs) in UI/UX design, synthesizing findings from 38 peer-reviewed studies published between 2022 and 2025. We identify key LLMs in use, including GPT-4, Gemini, and PaLM, and map their integration across the design lifecycle, from ideation to evaluation. Common practices include prompt engineering, human-in-the-loop workflows, and multimodal input. While LLMs are reshaping design processes, challenges such as hallucination, prompt instability, and limited explainability persist. Our findings highlight LLMs as emerging collaborators in design, and we propose directions for the ethical, inclusive, and effective integration of these technologies.
\end{abstract}

\begin{highlights} 
    \item This study is the first systematic literature review focused specifically on the integration of large language models (LLMs) into UI/UX design workflows.
    \item We identify key LLM applications, such as GPT-4 and Gemini, across the full design lifecycle, from ideation and prototyping to evaluation and refinement.
        \item Our review synthesizes best practices for LLM integration in design, while also addressing challenges including hallucinations, prompt instability, and limited explainability.
\end{highlights}



\maketitle

\section{Introduction}\label{sec1}

User Interface (UI) and User Experience (UX) design are foundational components of the software development lifecycle, playing a very important role in shaping how users perceive, interact with, and derive value from digital products. UI design encompasses the visual and interactive elements of a system, including layout, typography, and on-screen components. In contrast, UX design encompasses the broader user journey, including the emotions, perceptions, and behaviors that emerge before, during, and after interaction with a product \cite{kim2016technology}.

The quality of UI/UX design is a decisive factor in product success and user retention. Research consistently shows that poor UI/UX can drive users to abandon products altogether \cite{bharath2023leader, soegaard_bad_ux_2023}. Industry studies reinforce this: Forrester Research reports that well-designed UIs can improve website conversion rates by up to 200\%, and refined UX can lead to conversion gains of up to 400\% \cite{majumder2025influence}. Organizations that invest in UX design report measurable improvements in customer satisfaction, retention, and revenue \cite{majumder2025influence}. Some studies estimate that every dollar invested in UX returns as much as \$100 in value \cite{seattlenewmedia_nodate_uximportance}. Good UX is thus not just a matter of usability; it is a strategic business imperative \cite{alos2020importance}.

The role of the UI/UX designer extends from user research and behavioral analysis to wireframing, prototyping, and high-fidelity design. Designers synthesize user insights, structure information architecture, and iterate based on usability feedback \cite{chatterjee2017effects}. This multifaceted role requires balancing technical feasibility, business goals, and user needs, positioning designers as researchers, systems thinkers, and storytellers \cite{sonwalkar2019evolution}.

Despite the creative and user-centered nature of the design process, it remains time-intensive and cognitively demanding. Designers must simultaneously frame problems and craft effective solutions \cite{yang2017role}. As complexity increases, driven by tighter iteration cycles, stakeholder demands, and agile workflows, so does the risk of cognitive overload, burnout, and technostress \cite{douglas2023uxstress}. One survey found that over 90\% of UX professionals had experienced symptoms of burnout \cite{wojnarowska2020burnout}. Emerging technologies such as virtual reality (VR), augmented reality (AR), and the growing emphasis on accessibility standards (e.g., WCAG) introduce additional layers of complexity \cite{webaim2022million}. The need to master new tools and workflows further contributes to technostress, a phenomenon in which continuous adaptation to rapidly evolving technologies negatively affects performance and well-being \cite{kim2016technology}.

Artificial intelligence (AI) has long been viewed as a tool to ease these pressures. Even before the emergence of generative AI, researchers explored how AI could enhance various stages of the design lifecycle. AI has been applied to user context understanding \cite{yuan2023user, suleri2019eve, salminen2019design}, solution generation \cite{silva-rodriguez2020classifying, gomes2018artificial, sun2020developing, pandian2020blackbox, gardey2022predicting, duan2020optimizing}, design evaluation \cite{wallach2020beyond, swearngin2019modeling, yang2020measuring, zhou2020artificial}, and interface development \cite{dave2021survey, beltramelli2018pix2code, chen2018ui, latipova2019artificial, souza2020recent, moran2020machine}. For example, AI systems have shown potential in streamlining design tasks, reducing development time and cost \cite{oh2018lead}, and enabling adaptive interfaces responsive to evolving user needs \cite{johnston2019framework}.

The recent rise of generative AI, especially large language models (LLMs), introduces transformative possibilities for UI/UX design. LLMs have demonstrated advanced reasoning capabilities across a range of domains \cite{huang2023towards, plaat2024reasoning}, and the field continues to evolve rapidly \cite{zhao2023survey, zhou2023comprehensive, liu2023pretrain, dong2022survey}. According to the 2024 Stack Overflow Developer Survey, over 76\% of developers are currently using or planning to use AI tools in their workflows \cite{StackOverflow2024}. While LLMs are now widely adopted in software engineering tasks such as requirements elicitation \cite{ali2024large}, software testing \cite{wang2024software}, code completion \cite{husein2025large}, and deployment \cite{hou2024large}, their integration into UI/UX design remains comparatively under-explored in a structured and comprehensive manner.

This study seeks to fill that gap. By conducting a systematic literature review (SLR), we investigate how LLMs are currently used within UI/UX workflows, what best practices have emerged, and what challenges and risks practitioners face. Our contributions are as follows:

\begin{itemize}
    \item Identification and analysis of the open-source and proprietary Large Language Models (LLMs) currently employed by researchers and designers, including detailed insights into their applications within UI/UX design workflows.
    
    \item Compilation and discussion of best practices identified by researchers for effectively integrating LLMs into UI/UX design processes, aimed at maximizing benefits and improving workflow efficiency.
    
    \item Examination of the primary challenges and potential risks faced by researchers and practitioners when incorporating LLMs into UI/UX design, guiding in mitigating these challenges.
\end{itemize}

The remainder of this article is organized as follows: Section~\ref{related-surveys} reviews prior survey studies related to our topic. Section~\ref{methods} outlines the methodology of our systematic literature review, detailing the research questions, search strategy, screening and selection procedures, data extraction process, and the potential threats to validity. Section~\ref{results_discussion} presents the findings of our review, in which each subsection answers a particular research question. Finally, Section~\ref{conclusion} concludes the study and outlines directions for future research.

\section{Related Surveys} \label{related-surveys}

Several survey studies have explored the integration of artificial intelligence (AI) into UI/UX design workflows. While these efforts provide a valuable foundation for understanding AI-assisted design, they primarily address AI and machine learning (ML) in general and do not focus specifically on the role of Large Language Models (LLMs). To the best of our knowledge, this work represents the first systematic review dedicated to examining how LLMs are applied across the UI/UX design lifecycle. In this section, we summarize relevant prior surveys to contextualize the broader landscape of AI in design.

Stige et al.~\cite{stige2023ai} conducted a mapping study of AI applications in the UX design process, analyzing 46 articles. Their review focused on AI tools that supplement or replace designer tasks, with the majority of use cases concentrated in design solution generation (35\%) and automated development (31\%). Other applications included design evaluation, user context understanding, and prototyping. However, the authors noted a lack of clarity around which parts of the design process should be automated and emphasized concerns about AI diminishing designers’ creative autonomy. They concluded with a call for further investigation into the potential of tools like ChatGPT to support digital design processes.

Shi et al.~\cite{shi2023understanding} took a collaboration-centric view, analyzing 93 studies from the ACM Digital Library to explore how AI and designers interact. They identified two primary dynamics: AI augmenting designers by assisting with user need discovery, ideation, and layout generation; and designers enhancing AI through dataset labeling and alignment with creative goals. Key datasets used in these interactions include Rico~\cite{deka2017rico} and VINS~\cite{bunian2021vins}. The study introduced five dimensions of human-AI interaction, \textit{scope}, \textit{access}, \textit{agency}, \textit{flexibility}, and \textit{visibility}, and emphasized the need for explainable and adaptable AI tools that empower rather than replace designers.

Lu et al.~\cite{lu2024ai} analyzed 359 studies using the Human-Centered AI (HCAI) framework and the Double Diamond model (Discover, Define, Develop, Deliver) to categorize AI contributions to UX. They found that AI's integration remains limited, particularly due to its inability to support empathy-driven design and complex multi-screen workflows. Only 24.3\% of studies incorporated user-centered research methodologies, highlighting a significant imbalance between technology-driven and human-centered approaches.

Surveys by Bertao et al.~\cite{bertao2021aiuxui} and Chaudhry et al.~\cite{chaudhry2024aiux} examined industry perspectives. Bertao et al. surveyed 123 Brazilian UX/UI professionals and found limited adoption of AI, with most practitioners using it for automation and efficiency rather than creative tasks. Chaudhry et al., through a survey of 73 professionals, identified barriers to adoption including ethical concerns, lack of explainability, and fears of job displacement. Both studies highlighted the importance of AI literacy training and the need for stronger human oversight.

Xu et al.~\cite{xu2023aiux} approached the topic through the lens of Human-Computer Interaction (HCI), categorizing AI’s role as an assistant, collaborator, or facilitator. They highlighted explainability, user trust, and ethical risks as key challenges. Abbas et al.~\cite{abbas2022uxml}, focusing on ML adoption in UX design, followed the PRISMA framework to review 18 studies. Their findings emphasized the limited technical expertise of designers in ML, challenges in cross-disciplinary collaboration, and a lack of accessible AI-powered tools.

While these studies provide a broad overview of AI’s role in UI/UX workflows, they do not specifically address the use of generative LLMs, such as GPT-4, Bard, or Claude, in tasks like ideation, prototyping, content generation, or usability evaluation. Few if any of these surveys systematically analyze which LLMs are being used, how they are integrated into design workflows, or what best practices and risks are emerging in real-world applications. This review aims to fill that gap by conducting the first systematic literature review focused specifically on the role of LLMs in UI/UX design. We synthesize the current state of research, identify best practices for LLM integration, and highlight ongoing challenges and risks that need to be addressed in future design workflows.
\begin{figure*}
\centering 
\includegraphics[width=0.6\textwidth]{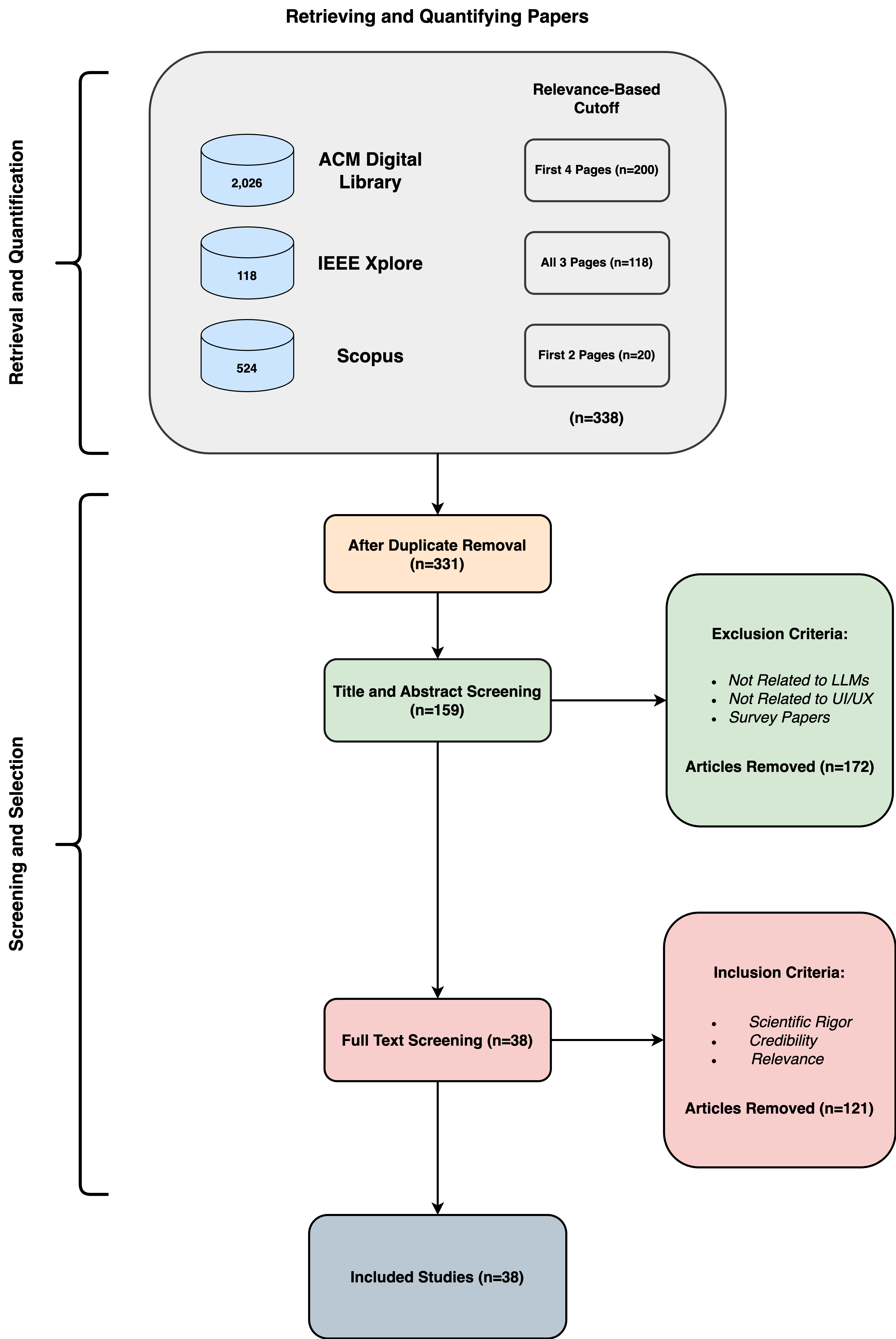}
\caption{PRISMA flow diagram illustrating the systematic selection process of studies on the use of large language models (LLMs) in UI/UX design. The diagram shows the initial search results across three academic databases (ACM Digital Library, IEEE Xplore, and Scopus), the application of relevance-based cutoff, duplicate removal, title and abstract screening based on defined exclusion criteria, and full-text screening using inclusion criteria, resulting in the final set of included studies.}
\label{fig:slr-process}
\end{figure*} 

\section{Methodology} \label{methods}
In this study, we aim to systematically investigate the state-of-the-art concerning the application of Large Language Models (LLMs) within UI/UX design workflows. We perform a systematic review following the mapping study methodology \cite{petersen2015guidelines}. The review comprehensively analyzes existing literature to identify and classify the types of LLMs employed, best practices for their integration into UI/UX workflows, and the associated challenges and risks. Our review adheres to the guidelines proposed by Kitchenham et al.~\cite{kitchenham2009systematic} and involves the following phases:
\begin{enumerate}
    \item Defining research questions
    \item Developing the search strategy
    \item Identifying relevant primary studies
    \item Extracting relevant data from selected studies
    \item Assessing threats to validity
\end{enumerate}

\subsection{Research Questions}
To conduct a comprehensive and detailed analysis, the overarching research objective is divided into three focused research questions (RQs), outlined in Table~\ref{tab:research-questions}. These RQs guide the systematic review by clearly defining its scope, structure, and investigative path, enabling readers to effectively follow and interpret the findings of this study.

\begin{table*}[h!]
\centering
\caption{Defined research questions for the systematic literature review}
\label{tab:research-questions}
\begin{tabular}{p{1cm}p{12cm}}
\hline
\textbf{ID} & \textbf{Research Question (RQ)} \\ \hline
RQ1 & What open-source and proprietary Large Language Models (LLMs) are currently utilized/used to assist in the UI/UX design tasks, and how are these LLMs being integrated into UI/UX design workflows? \\ \hline
RQ2 & What are the best practices identified in the literature for effectively integrating LLMs into UI/UX design processes? \\ \hline
RQ3 & What are the limitations and challenges identified by researchers associated with incorporating LLMs into UI/UX design tasks or workflows? \\ \hline
\end{tabular}
\end{table*}

\subsection{Search Strategy}
We conducted a systematic literature search using three academic databases: ACM Digital Library, IEEE Xplore, and Scopus. The final Boolean search query used for retrieving relevant studies is presented in Table~\ref{tab:keywords-query}. The initial number of papers identified from each database, along with subsequent screening and selection steps, is illustrated in Figure~\ref{fig:slr-process}. To ensure relevance to recent advancements, the search was limited to papers published between 2022 and 2025, reflecting the widespread commercial availability and adoption of Large Language Models such as ChatGPT since late 2022. 

\subsubsection{Relevance-based Cutoff Strategy}

\begin{figure*}
\centering 
\includegraphics[width=0.4\linewidth]{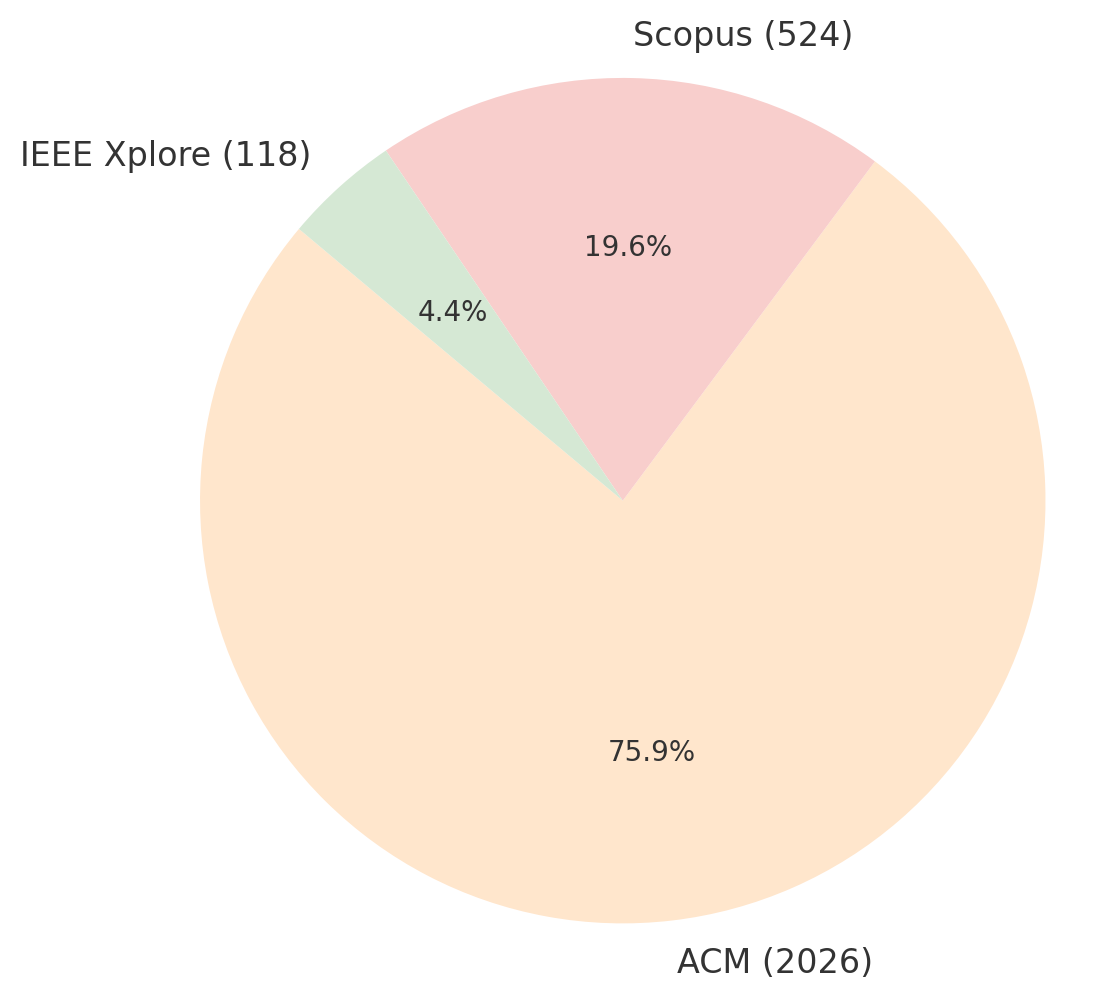}
\caption{Distribution of fetched publications using our search query across ACM, Scopus, and IEEE Xplore databases.}
\label{fig:dist-articles}
\end{figure*}

Our initial search yielded 2,668 papers across three databases (Figure \ref{fig:dist-articles}), with ACM returning the highest number. To manage the volume effectively, we employed a relevance-based cutoff strategy tailored to each database. We reviewed the titles of the top 10 articles on each results page (each page contains 20 articles). If at least 5 of these 10 titles aligned closely with our research focus, e.g., including terms like design, UI/UX, AI, LLMs, or chatbots, we considered the entire page relevant and included all 20 articles. Once a page showed a marked drop in relevance (i.e., fewer than 5 of the top 10 titles met the criteria), we discontinued review of subsequent pages in that database.

This method does not guarantee that every included paper was relevant, but it allowed for a pragmatic balance between thoroughness and feasibility. For instance, while ACM Digital Library returned 2,026 papers, most relevant studies were concentrated within the first 10 pages. Pages beyond this point were excluded due to declining relevance. IEEE Xplore returned only three pages, all of which were included. Scopus yielded relevant results primarily on the first two pages. Springer results were excluded entirely due to a lack of relevance. Ultimately, our relevance-based filtering led to a final dataset of 338 papers across the three selected databases for further analysis.

\begin{table*}[h!]
\centering
\caption{Keyword groups and the final boolean search query}
\label{tab:keywords-query}
\begin{tabular}{p{3cm}|p{10cm}}
\hline
\textbf{Groups} & \textbf{Keywords} \\ \hline
LLM-related keywords & Large Language Models, LLMs, GPT, ChatGPT, Chatbot \\ \hline
UI/UX-related keywords & UI Design, UX Design, UI/UX Design, User Interface Design, User Experience Design, Frontend Design, Frontend Development, Web Development, Web Design, Prototyping, Personas \\ \hline
\textbf{Final Boolean Query} & ("Large Language Models" OR "LLMs" OR  "GPT" OR "ChatGPT" OR "Chatbot") AND ("UI Design" OR "UX Design" OR "UI/UX Design" OR "User Interface Design" OR "User Experience Design" OR "Frontend Development" OR "Web Development" OR "Web Design" OR "Frontend Design" OR "Prototyping" OR "Personas") \\ \hline
\end{tabular}
\end{table*}

\subsection{Screening and Selection}
Given the initial set of 338 papers, we began by removing duplicates using Zotero reference manager, which identified and eliminated a total of seven duplicates. Additionally, one non-English article was identified and excluded. Subsequently, we applied the predefined exclusion criteria outlined in Table \ref{tab:inclusion-exclusion}. Following this, we carefully evaluated the remaining papers, selecting primary studies based on relevance, scientific rigor, and credibility. The resulting final set of studies was then advanced for detailed analysis.

\begin{table}[h]
\centering
\caption{Inclusion and Exclusion Criteria}
\label{tab:inclusion-exclusion}
\begin{tabular}{c|c}

\textbf{Exclusion Criteria} &
\parbox{3.2cm}{%
    \centering
    Not Related to LLMs,\\
    Not Related to UI/UX,\\
    \emph{Survey Papers / Book Chapters / Reviews}
} \\
\hline
\textbf{Inclusion Criteria} &
\parbox{3.2cm}{%
    \centering
    Scientific rigor,\\
    Relevance, and\\
    Credibility
} \\

\end{tabular}

\end{table}

\subsubsection{Exclusion Criteria}
Papers were excluded if they did not directly relate to Large Language Models (LLMs) or UI/UX processes (such as prototyping, wireframing, content design, user flow optimization, usability testing, etc). Additionally, survey papers, book chapters, and literature reviews were removed, although these represented only a small fraction (fewer than 10 survey papers and one or two book chapters or reviews).

During the initial abstract screening, many papers were excluded for lacking clear relevance to either LLMs or UI/UX. At this preliminary stage, papers mentioning LLMs and UI/UX, even superficially, were retained. However, we have not yet verified if the studies specifically used LLMs to support UI/UX workflows. Thus, some studies discussing the integration of chatbots or addressing general UX/UI aspects without explicitly employing LLMs in the design or enhancement of user experiences might remain. In this initial screening stage, we removed a total of 172 studies and were left with 159.

\begin{figure*}
\centering 
\includegraphics[width=0.6\linewidth]{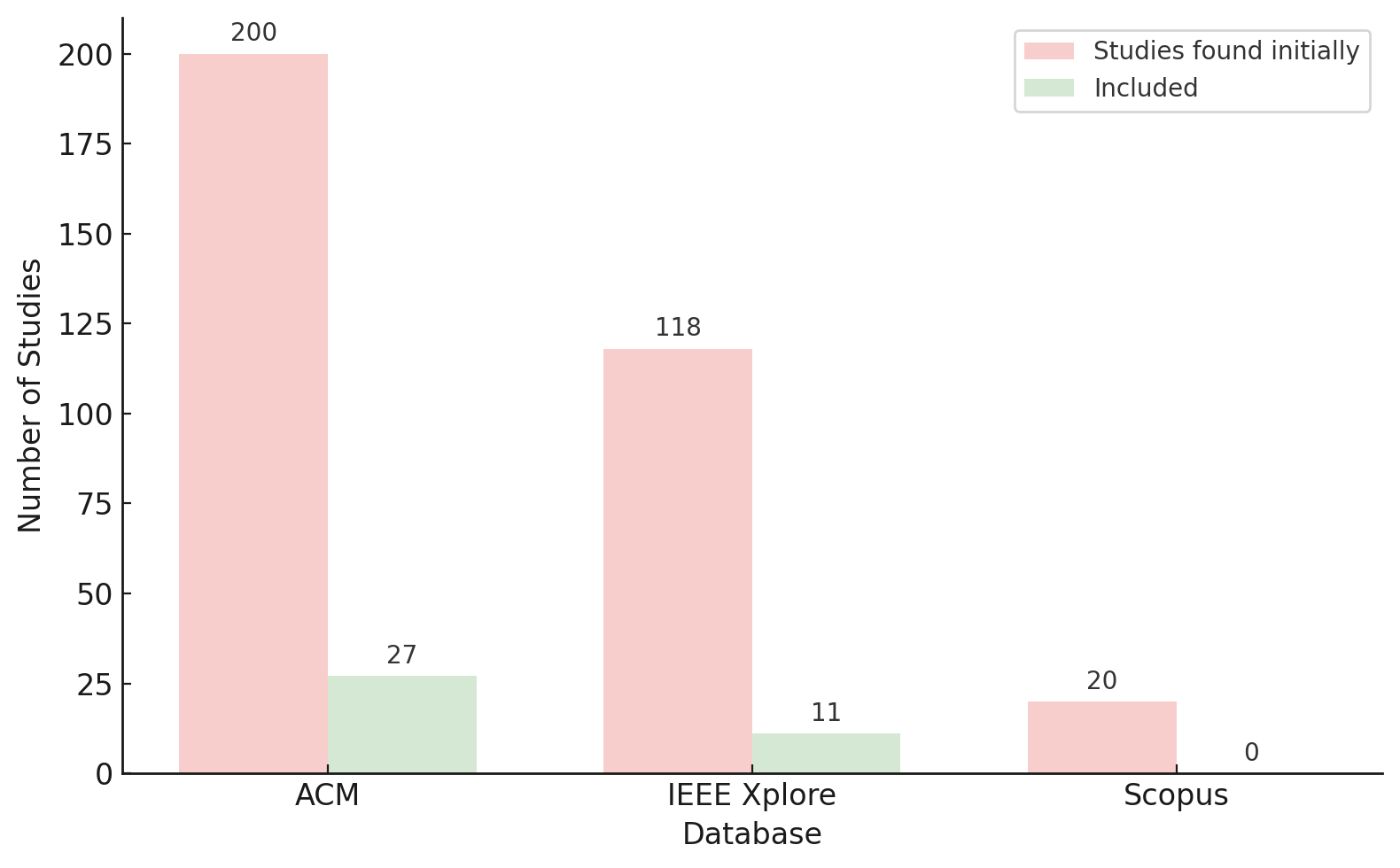}
\caption{Number of studies retrieved and included from each digital library. The majority of included studies were sourced from ACM and IEEE Xplore, while no studies from Scopus met the exclusion and inclusion criteria.}
\label{fig:included}
\end{figure*}

\subsubsection{Inclusion Criteria}
We conducted a comprehensive full-text screening of each article based on three key criteria. First, we assessed credibility by confirming that each article was published in a reputable, peer-reviewed journal or recognized conference proceedings. Second, we evaluated relevance by determining whether the study directly addressed our research questions, specifically focusing on the use of Large Language Models (LLMs) within UI/UX workflows, the generalizability of best practices to other UI/UX contexts, and insights into potential challenges or risks associated with integrating LLMs in UI/UX design. Third, we scrutinized the scientific rigor of each study, ensuring that the methodology was sound and transparent, and that results were clearly articulated, logically presented, and thoroughly supported by evidence. 

Most articles were published in the ACM Digital Library, a reputable source, thereby meeting our credibility criterion. However, a few lacked scientific rigor; for example, some were too brief or did not provide enough methodological detail to enable reproducibility. More importantly, relevance emerged as the principal shortcoming. While many papers did employ Large Language Models (LLMs), they focused on narrow, domain-specific applications and saw how the integration of LLMs could improve a specific process, rather than broader UI/UX design workflows. For instance, some studies described how LLMs were used to develop an educational app (which happens to contain UI/UX) but did not examine how or why these models improved the UI/UX design process, nor did they offer insights generalizable to other design projects. Hence, we did not feel that these studies yielded any lessons applicable to a wider UI/UX context, therefore, we decided not to include such studies.

Although the keyword "Personas" was initially included in our search query, we later recognized that studies focused primarily on the generation, evaluation, or diversity of personas using LLMs do not align closely with the core objectives of this systematic literature review. While personas remain a valuable tool in UI/UX design, the research questions guiding this review emphasize the practical integration of LLMs into broader design workflows, as well as the identification of generalizable best practices for such integration. In contrast, persona-centered studies tend to focus on the representation of user archetypes, often in isolation from key design activities such as prototyping, wireframing, user testing, or iterative development. Such studies do not meet our inclusion criteria, which require (1) clear documentation of LLM integration within design workflows (RQ1) and (2) the presentation of actionable, reusable best practices (RQ2). Nonetheless, the inclusion of the ``Personas'' keyword did not compromise the relevance of our results, as we applied a relevance-based cut-off strategy to determine when to cease retrieval, ensuring that only pertinent studies were included.


 \begin{figure*}
\centering 
\includegraphics[width=0.6\linewidth]{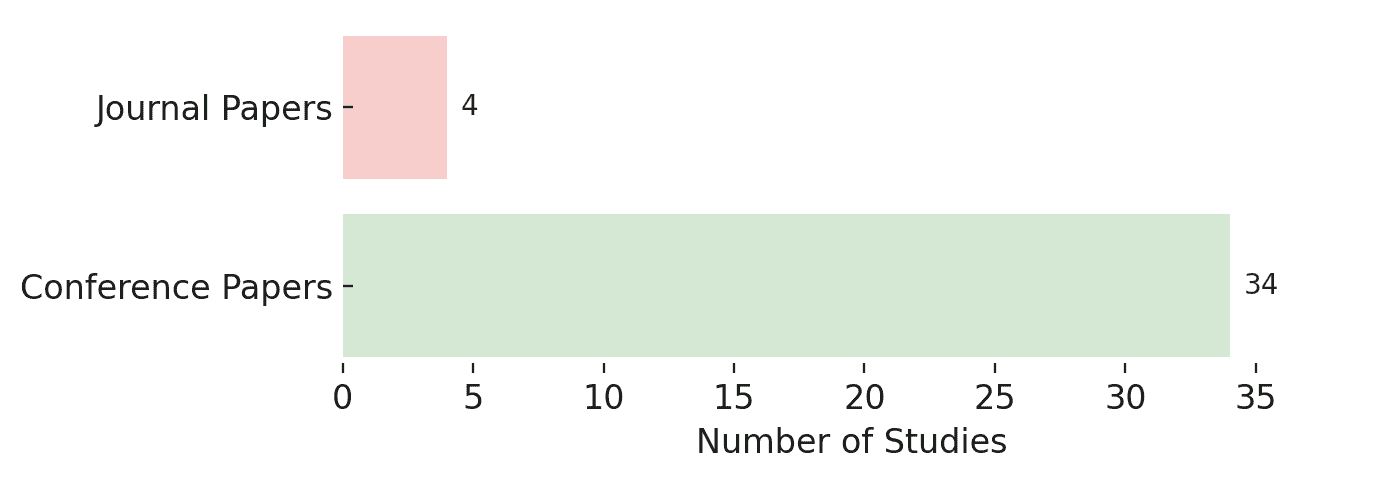}
\caption{Distribution of journal vs. conference papers included for analysis.}
\label{fig:journal-conference}
\end{figure*} 

\subsubsection{Included Studies}
After the full text screening, satisfying all three criteria as stated in Table \ref{tab:inclusion-exclusion}, a total of 38 studies were included. Since utilizing LLMs in UI/UX design is an emerging field, it was expected that a final pool of relevant, high-quality studies would be smaller. While the number may appear modest, the inclusion criteria were deliberately strict to ensure each study addressed both the use of LLMs in UI/UX workflows and provided transferable best practices. Given the novelty and specificity of the research topic, this number is both expected and justified. The number of studies that we initially retrieved and included from each database is illustrated in Figure \ref{fig:included}. Majority of the included studies were from ACM, a total of 27 out of 38 studies, this is to be expected, since many ACM venues actively encourage cross-cutting work across AI, design, education, creativity, and human factors, making them more likely to publish relevant studies where LLMs are evaluated in real-world or design-specific workflows. In contrast, many studies retrieved from IEEE Xplore and Scopus focused primarily on algorithmic aspects of LLMs without addressing design integration or transferable best practices. Nevertheless, 11 out of 38 were highly relevant from IEEE Xplore, and none were included from Scopus.

\begin{figure*}
\centering 
\includegraphics[width=1\linewidth]{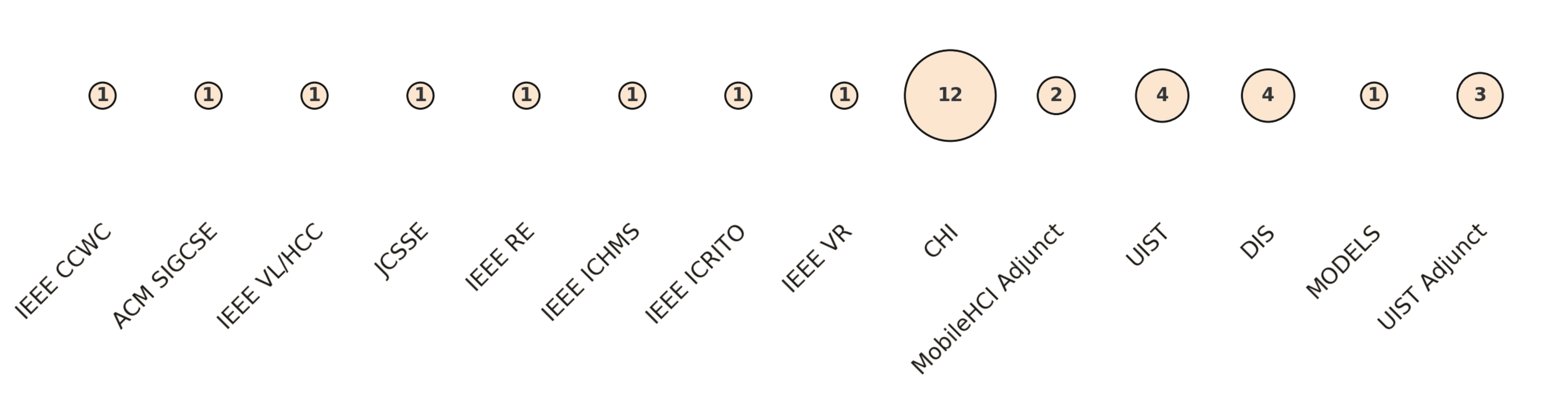}
\caption{Distribution of included studies across ACM and IEEE conferences. The bubble sizes represent the number of publications per venue.}
\label{fig:bubble-conference}
\end{figure*}

Another notable observation is that the majority of the included studies were conference papers rather than journal articles, as shown in Figure \ref{fig:journal-conference}. This aligns with the nature of the research area, as conferences, particularly in HCI and AI-related fields, tend to be the primary venues for publishing cutting-edge, exploratory, and interdisciplinary work. Conferences like CHI, DIS, and IUI often serve as rapid dissemination platforms for novel tools, user studies, and design innovations, which match the scope and focus of this review. Moreover, all 4 journal papers were from IEEE. Figure \ref{fig:bubble-conference}, shows the distribution of the included conference studies across various ACM and IEEE conferences. This visualization highlights the most frequent venues, Conference on Human Factors in Computing Systems (CHI) being by far the most frequent, guiding researchers toward optimal publication outlets for studies aligned with the theme of this systematic review.

\begin{figure*}
\centering 
\includegraphics[width=0.6\linewidth]{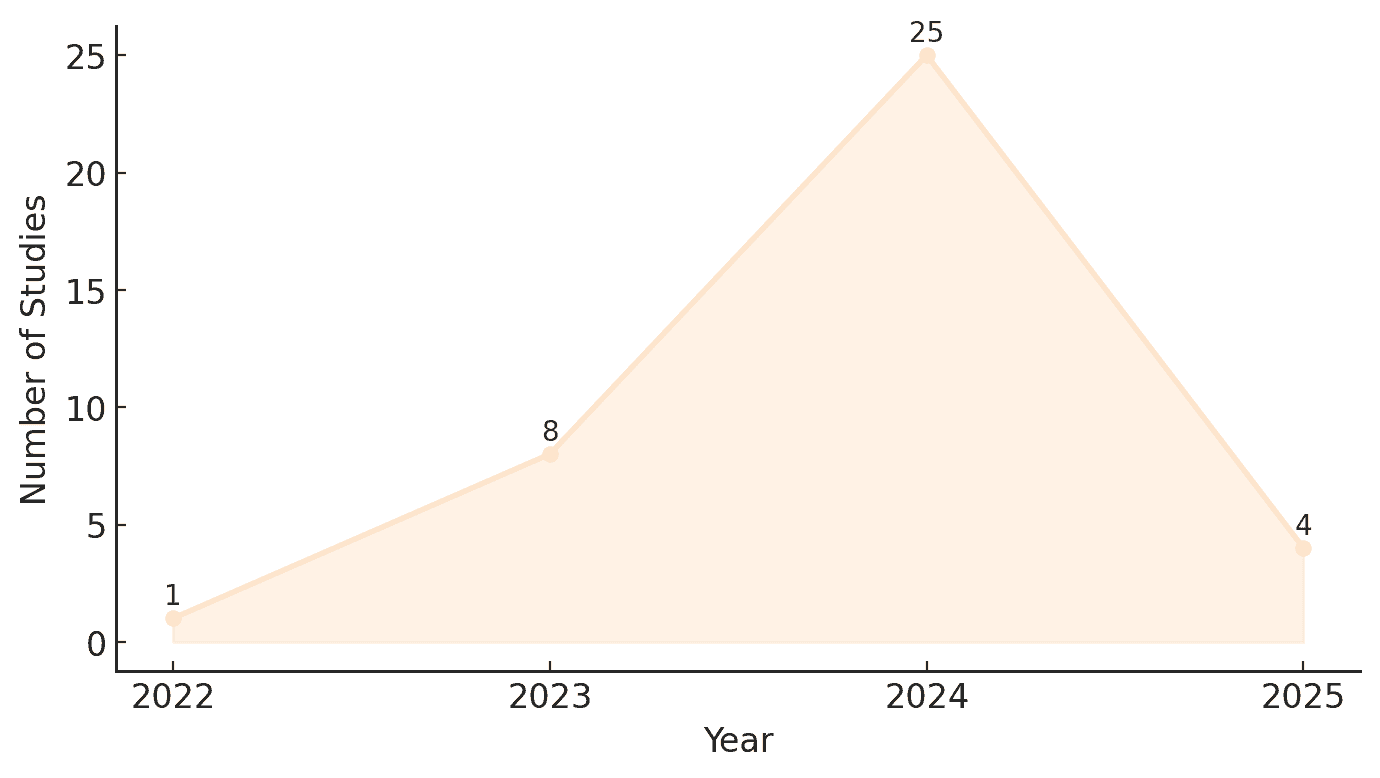}
\caption{Distribution of included studies by publication year. The number of relevant studies has grown rapidly since 2022, with a notable peak in 2024, indicating increasing academic attention to the role of LLMs in UI/UX design workflows.}
\label{fig:pub-year}
\end{figure*}

We can also see that many of these studies have been recently published, as shown in Figure \ref{fig:pub-year}. The trend in publication years indicates a rapidly growing research interest in the integration of LLMs into UI/UX design workflows. From just 1 study in 2022, the number increased significantly to 8 in 2023, peaking at 25 studies in 2024. This sharp rise suggests that the topic has recently gained momentum, likely driven by advances in accessible LLMs (e.g., GPT-4) and their integration into design tools. At the time of writing this, the fact that we are only 3 months in and we already have 4 studies, shows that the upward trajectory appears to be continuing.

\begin{table*}[htbp]
\centering
\caption{Summary of studies considered (sorted by citations, descending)}
\label{tab:study-summary}
\begin{tabular}{cccccc}
\hline
\textbf{Study} & \textbf{Year} & \textbf{Type} & \textbf{Publisher} & \textbf{Concise Topic} & \textbf{Citations} \\
\hline
\cite{jiang2022promptmaker} & 2022 & Conference Paper & ACM & Prompt-based Prototyping with LLMs & 65 \\
\cite{wang2023conversationalui} & 2023 & Conference Paper & ACM & Mobile UI Interaction via LLMs & 63 \\
\cite{bilgram2023aiinnovation} & 2023 & Journal Article & IEEE & AI-Augmented Prototyping Methods & 59 \\
\cite{zamfirescu2023herding} & 2023 & Conference Paper & ACM & Designing Chatbots by Prompting GPT-3 & 30 \\
\cite{kim2023cells} & 2023 & Conference Paper & ACM & Object-Oriented Interaction with LLMs & 21 \\
\cite{petridis2023promptinfuser} & 2023 & Conference Paper & ACM & UI Mock-up Animation with LLMs & 19 \\
\cite{tian2025chartgpt} & 2025 & Journal Article & IEEE & Chart Generation from Natural Language & 12 \\
\cite{masson2024directgpt} & 2024 & Conference Paper & ACM & Direct Manipulation Interfaces for LLMs & 8 \\
\cite{leiser2024hill} & 2024 & Conference Paper & ACM & Identifying Hallucinations in LLMs & 8 \\
\cite{delatorre2024llmr} & 2024 & Conference Paper & ACM & Real-Time Prompting of Interactive Worlds & 7 \\
\cite{takaffoli2024genaiux} & 2024 & Conference Paper & ACM & GenAI Use in UX Practice & 6 \\
\cite{wang2024farsight} & 2024 & Conference Paper & ACM & Responsible AI in Prototyping & 5 \\
\cite{duan2024uifeedback} & 2024 & Conference Paper & ACM & Auto Feedback on UI Mockups via LLMs & 5 \\
\cite{vaithilingam2024dynavis} & 2024 & Conference Paper & ACM & Synthesized UI Widgets for Visualization & 4 \\
\cite{giunchi2024dreamcodevr} & 2024 & Conference Paper & IEEE & Behavior Design in VR via AI & 3 \\
\cite{duan2023uifeedback} & 2023 & Conference Paper & ACM & UI Design Feedback with LLMs & 3 \\
\cite{du2023visualblocks} & 2023 & Conference Paper & ACM & Visual Prototyping of AI Pipelines & 3 \\
\cite{xiang2024simuser} & 2024 & Conference Paper & ACM & Simulated User Interaction for Usability & 2 \\
\cite{song2024visiontasker} & 2024 & Conference Paper & ACM & Task Automation via Vision UI and LLMs & 2 \\
\cite{petridis2024promptinfuser} & 2024 & Conference Paper & ACM & Coupling AI and UI Design & 2 \\
\cite{liu2023visualcaptions} & 2023 & Conference Paper & ACM & Real-Time Visual Communication with LLMs & 2 \\
\cite{zhou2024instructpipe} & 2024 & Conference Paper & ACM & Multi-modal Pipelines with LLMs & 1 \\
\cite{shaer2024integrating} & 2024 & Journal Article & IEEE & GenAI in Tangible Interaction Courses & 1 \\
\cite{wu2024uiclip} & 2024 & Conference Paper & ACM & UI Design Assessment via Data-Driven Model & 1 \\
\cite{patel2025starrystudioai} & 2025 & Conference Paper & IEEE & Generative AI and RAG for UI Design & 0 \\
\cite{wei2025aiui} & 2025 & Journal Article & IEEE & AI-Inspired UI Design & 0 \\
\cite{aljedaani2025accessibility} & 2025 & Conference Paper & ACM & LLMs for Software Accessibility & 0 \\
\cite{petridis2024insitu} & 2024 & Conference Paper & IEEE & Multimodal Prompts in Mobile Prototyping & 0 \\
\cite{kolthoff2024interlinking} & 2024 & Conference Paper & IEEE & Linking User Stories and GUI Prototypes & 0 \\
\cite{oruche2024holistic} & 2024 & Conference Paper & IEEE & Multi-layered Design for HCD Systems & 0 \\
\cite{singhal2024largeaction} & 2024 & Conference Paper & IEEE & Next-Gen Web and App Engagement & 0 \\
\cite{liu2024usability} & 2024 & Conference Paper & IEEE & Usability of AI-Enhanced Video Apps & 0 \\
\cite{zhang2024designwatch} & 2024 & Conference Paper & ACM & User Operation Analysis in Mobile Apps & 0 \\
\cite{wang2024aigc} & 2024 & Conference Paper & ACM & AIGC’s Role in UX Collaboration & 0 \\
\cite{sasaki2024geofence} & 2024 & Conference Paper & ACM & Geofencing City Walks with LLMs & 0 \\
\cite{lu2024aiisnotenough} & 2024 & Conference Paper & ACM & AI and Heuristics in UI Linting & 0 \\
\cite{duan2024uicrit} & 2024 & Conference Paper & ACM & UI Evaluation Using Critique Dataset & 0 \\
\cite{benchaaben2024graphicalsyntax} & 2024 & Conference Paper & ACM & Generating Tailored Graphical Syntax & 0 \\
\hline
\end{tabular}
\end{table*}

\subsection{Data Extraction}
To systematically extract and organize relevant data from all 38 included studies, we utilized a structured Microsoft Excel spreadsheet. This spreadsheet comprised nine distinct fields, each designed to capture critical information pertinent to our analysis, as outlined in Table \ref{tab:extraction_fields}. The \textit{Source} field records the URL link to each study to ensure traceability. We included a \textit{Citations} field to note the number of citations per study, serving as a proxy for its academic impact. The \textit{Objectives} field summarizes the key goals or research questions addressed in each study. We documented the \textit{LLMs Used}, listing the specific large language models employed, while the \textit{Datasets} field notes any datasets utilized for training, evaluation, or fine-tuning, if any. To assess performance, we included an \textit{Evaluation Metrics} field, capturing the quantitative or qualitative measures reported by the authors. To identify actionable insights, we added a \textit{Best Practices} field, summarizing effective strategies or methodological recommendations for integrating LLMs into UI/UX tasks. Finally, the \textit{Limitations} field highlights challenges, constraints, or known issues associated with the application of LLMs in UI/UX domains.

\begin{table*}[h!]
\centering
\caption{Data extraction fields and their descriptions}
\label{tab:extraction_fields}
\begin{tabular}{@{}ll@{}}
\toprule
\textbf{Field} & \textbf{Description} \\
\midrule
Source & URL link of the study to ensure traceability and context. \\
Citations & Number of citations to indicate the study’s academic influence. \\
Objectives & Key goals, research questions, or hypotheses addressed in the study. \\
LLMs Used & Specific large language models utilized, such as GPT-3, GPT-4, Gemini, Claude, etc. \\
Datasets & Datasets used for training, evaluation, or analysis (if any). \\
Evaluation Metrics & Measures used to assess model performance or user interaction outcomes, quantitatively and/or qualitatively. \\
Best Practices & Effective strategies, workflows, or recommendations identified for applying LLMs in UI/UX contexts. \\
Limitations & Challenges, drawbacks, or limitations related to the use of LLMs in UI/UX tasks. \\
\bottomrule
\end{tabular}
\end{table*}

\subsection{Threats to Validity}
Despite our adherence to systematic review guidelines and rigorous screening procedures, several potential threats to validity exist that we feel merit discussion.

\subsubsection{Selection Bias}
To manage the scope of the review, we applied a relevance-based cutoff strategy during the search phase, where we ceased reviewing results from academic databases after observing a significant drop in topical relevance. While this approach improves efficiency, it introduces a risk of excluding relevant studies buried deeper in the search rankings. Additionally, results from the Scopus database were excluded entirely due to an initial lack of relevant findings, which may have inadvertently omitted valuable contributions not indexed by ACM or IEEE. 

\subsubsection{Publication Bias} The majority of included studies were conference papers, which, although timely and often innovative, may not always undergo the same level of peer-review scrutiny as journal articles. As such, the findings may lean toward exploratory or early-stage research. While this reflects the nascent and rapidly evolving nature of the LLM-UX integration field, it may affect the generalizability and maturity of the synthesized insights.

\subsubsection{Language Bias} Only English-language studies were considered in this review. This constraint may exclude relevant work published in other languages, potentially underrepresenting research from non-English-speaking regions where UX practices or LLM adoption may differ.

\subsubsection{Terminological Ambiguity} Although the Boolean search string was carefully crafted, the inclusion of broader terms such as “chatbot” and “personas” led to the retrieval of studies only tangentially related to our research questions. While these were excluded during full-text screening, their presence may have introduced initial noise and influenced the scope of the review. Moreover, certain terms (e.g., “prototyping”) may be interpreted differently across domains, complicating consistent filtering.

\begin{figure*}
\centering 
\includegraphics[width=1\linewidth]{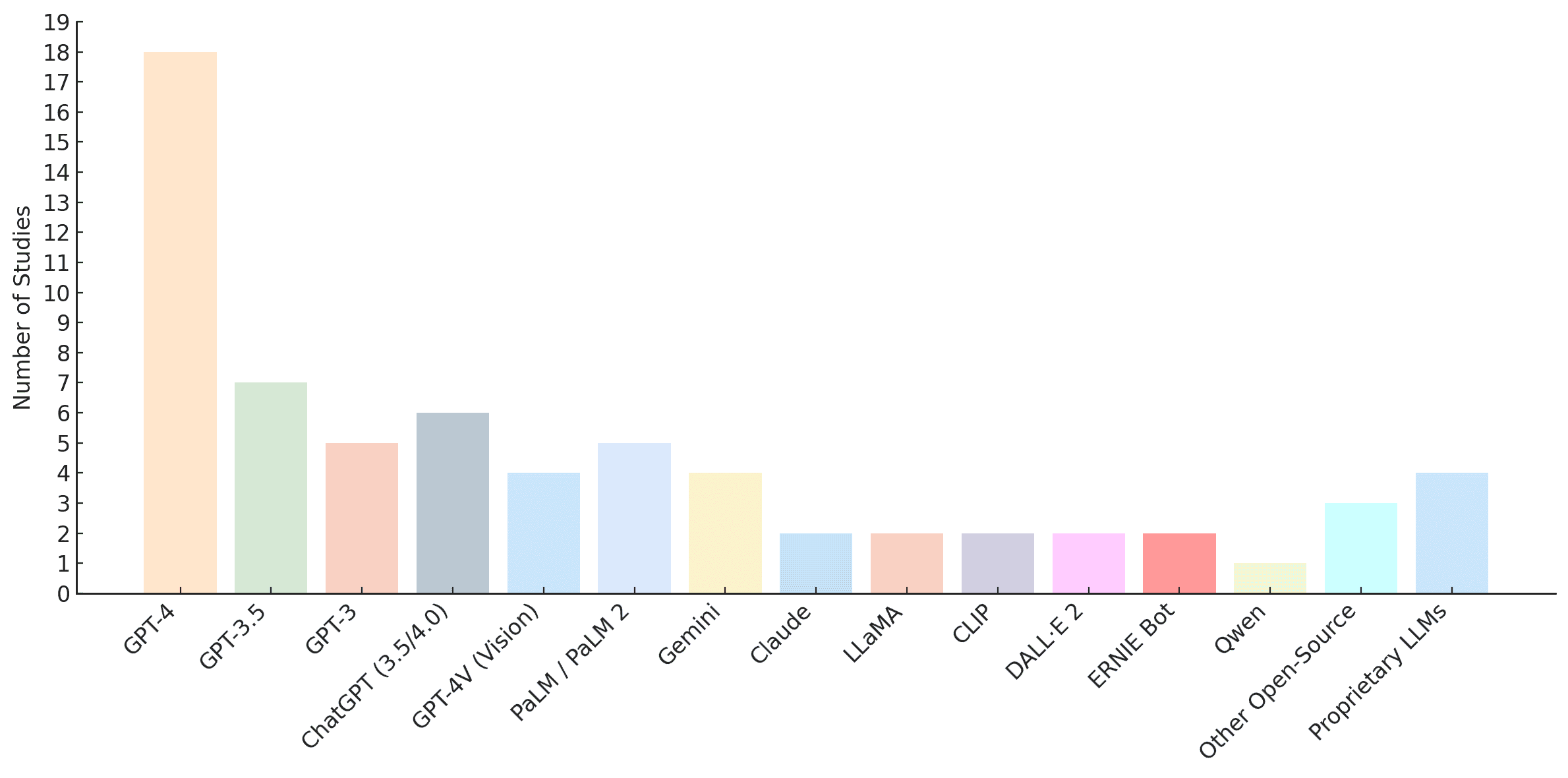}
\caption{Frequency of large language model (LLM) usage across reviewed UI/UX design studies.}
\label{fig:llms-used}
\end{figure*}

\begin{table*}[htbp]
\centering
\caption{Frequency of LLM usage in all reviewed studies.}
\begin{tabular}{l|l|l}
\hline
\textbf{Models} & \textbf{\# of Studies} & \textbf{Notes} \\
\hline
GPT-4 & 18 & Widely used for prompting, UI generation, reasoning, and multimodal tasks. \\
GPT-3.5 & 7 & Frequently accessed via ChatGPT; used in prototyping and co-creation. \\
GPT-3 & 5 & Applied in ideation, dataset generation, and early LLM workflows. \\
ChatGPT (3.5/4.0) & 6 & Used for the same tasks as GPT-4 and 3.5. \\
GPT-4V (Vision) & 4 & Used for multimodal UI interpretation and simulation. \\
PaLM / PaLM 2 & 5 & Used for generation, summarization, and vision-language tasks. \\
Gemini (1 / Pro / Vision) & 4 & Employed in structured code generation, screenshot-based UI feedback. \\
Claude (2 / 3) & 2 & Used as an alternative or in benchmarking. \\
LLaMA (2 / 3) & 2 & Used in fine-tuned or instruction-tuned variants for UI code output. \\
CLIP (fine-tuned) & 2 & Used for UI relevance and interpretation, sometimes with GPT support. \\
DALL·E 2 & 2 & Used for sketch and visual design generation, supporting LLM workflows. \\
ERNIE Bot (Baidu) & 2 & Applied in task planning and automation systems. \\
Qwen (e.g., 1.5-110B) & 1 & Benchmarked; not a primary model. \\
LLaVA, Falcon, Nous, etc. & 1 (each) & Evaluated or benchmarked, not core to implementations. \\
Unnamed Proprietary LLMs & 4+ & Used in internal tools (e.g., ARChat, Visual Captions, PromptMaker) \\
\hline
\end{tabular}
\label{tab:llm-usage}
\end{table*}

\section{Results and Discussion} \label{results_discussion}
This section presents a comprehensive overview and analysis of the findings from all 38 included studies. Each subsection is organized to address a specific research question, providing a structured and focused discussion of the results.

\subsection{\textit{RQ1: What open-source and proprietary Large Language Models (LLMs) are currently utilized/used to assist in the UI/UX design tasks, and how are these LLMs being integrated into UI/UX design workflows?}}

We will first answer the first part of this research question, namely \textit{``What open-source and proprietary Large Language Models (LLMs) are currently utilized/used to assist in the UI/UX design tasks''}. Across the surveyed literature, GPT-4 stands out as the most widely utilized large language model, as illustrated in the Figure \ref{fig:llms-used}, owing to its strong performance in structured UI generation, reasoning through Chain-of-Thought (CoT) prompting, and its support for multimodal inputs. Many studies rely directly on GPT-4 APIs, while others integrate the model through platforms like ChatGPT or specialized toolkits for prototyping, code generation, or user simulation. GPT-3.5 and GPT-3 are also commonly used; however, they increasingly serve as baseline or comparison models, reflecting the field’s evolution toward more capable systems.

Google’s PaLM and Gemini families are also gaining traction, particularly in studies involving multimodal tasks, screenshot-based analysis, and UI feedback generation. On the other hand, models like Claude (Anthropic) and LLaMA (Meta) appear more often as alternatives or in benchmarking roles, with limited direct usage in implementations. Notably, several studies adopt a model-agnostic design, allowing for interchangeable use of LLMs via API endpoints, although OpenAI’s models remain the most frequently evaluated in practice.

An important trend is the increased use of vision-language models, such as GPT-4V, Gemini Pro Vision, and PaLI, to interpret and generate UI components from visual inputs like screenshots or user interactions. Additionally, a number of studies rely on proprietary or internal LLMs, particularly in industrial or corporate research settings, where toolchains may mask the underlying model. These cases reflect the growing prevalence of LLM-as-a-service paradigms, where researchers interact with models through high-level APIs or plugins (e.g., Figma, Chrome extensions) without always disclosing model details. Finally, it’s worth noting that many studies utilize more than one LLM, either to compare outputs, enhance performance through ensemble methods, or balance latency and accuracy trade-offs across different system components. As illustrated in Figure \ref{fig:llms-used}, GPT-4 is the most commonly used LLM in UI/UX design research, appearing in 18 studies. While GPT-3.5 and GPT-3 still play important roles, their usage is increasingly limited to baseline comparisons. The use of models such as PaLM, Gemini, and GPT-4V also reflects a broader trend toward multimodal and design-specific applications. These findings are also summarized in the form of short notes in Table \ref{tab:llm-usage}.

\begin{figure*}
\centering 
\includegraphics[width=1\linewidth]{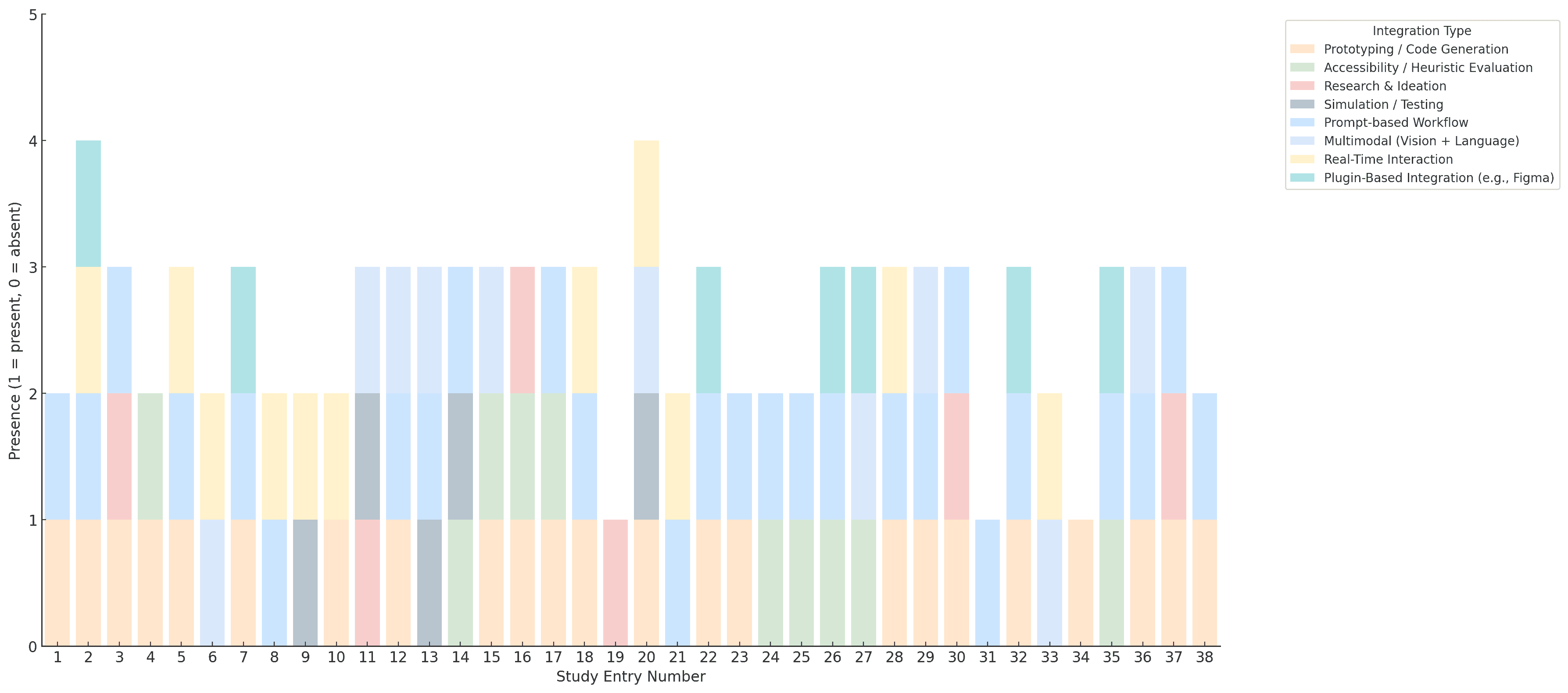}
\caption{Distribution of LLM integration types across reviewed UI/UX studies (see Appendix \ref{appendix}). Each bar represents a study, with stacked segments indicating the presence of specific LLM integration types.}
\label{fig:llm-integration-studies}
\end{figure*}

We now move on to the second part of this research question, namely \textit{``How are these LLMs being integrated into UI/UX design workflows''}. The integration of LLMs into UI/UX design workflows is marked by significant diversity, creativity, and a clear evolution toward human-AI collaboration and workflow augmentation. Across the 38 reviewed studies, we observe several prominent trends in how LLMs, both proprietary and open-source, are operationalized within real-world and experimental design contexts. 

\subsubsection{LLMs as Embedded Design Tools}
A strong trend we observed is the embedding of LLMs directly within existing design platforms like Figma, Unity, and Google Meet via plugins, extensions, or API interfaces.  Studies using tools such as PromptInfuser \cite{petridis2023promptinfuser, petridis2024promptinfuser}, MobileMaker \cite{petridis2024insitu}, LLMR \cite{delatorre2024llmr}, or DirectGPT \cite{masson2024directgpt} show that LLMs are often not standalone systems, but components within broader toolchains. The integration takes several forms. First, GPT-4 and Gemini are embedded into design tools to automate heuristic evaluations, generate HTML/CSS, or provide usability suggestions \cite{duan2024uifeedback, duan2024uicrit, duan2024uifeedback}. Second, LLMs are used inside tools like PromptInfuser or MobileMaker, enabling real-time prototyping through natural language commands and creating prototyping interfaces \cite{petridis2023promptinfuser, petridis2024promptinfuser, petridis2024insitu}. Third, systems like SimUser \cite{xiang2024simuser} and LLMR \cite{delatorre2024llmr} use LLMs as agents simulating both user behavior and app functionality, enabling heuristic evaluation without real users. 

Hence, plugin-based LLM integration allows designers to interact with LLMs using their habitual tools, including heuristic evaluations \cite{duan2023uifeedback, duan2024uifeedback}, UI code generation \cite{patel2025starrystudioai}, or integrating AI agents into mobile and XR design \cite{delatorre2024llmr, giunchi2024dreamcodevr}. This mode of integration allows workflow continuity, enabling rapid, in-context AI support without requiring designers to switch environments or tools. Therefore, the LLM is no longer peripheral but increasingly positioned as an integral, interactive component within mainstream UI/UX toolchains.

\subsubsection{Prompt-Based Interaction as a Core Paradigm}
Among the 38 reviewed studies, natural language prompting was a near-universal mode of interaction. More specifically, a dominant integration pattern we observed was \textit{prompt-based prototyping}, where LLMs are driven through structured natural language commands. Rather than retraining or fine-tuning models, most systems rely on Zero-shot / Few-shot prompting \cite{kolthoff2024interlinking, duan2024uicrit, wang2023conversationalui}, Chain-of-Thought (CoT) strategies for task decomposition \cite{tian2025chartgpt, xiang2024simuser, oruche2024holistic}, Retrieval-Augmented Generation (RAG) for domain-specific grounding \cite{patel2025starrystudioai, lu2024aiisnotenough}. Prompting supports:

\begin{itemize}
    \item Code generation (HTML/CSS, JavaScript, SVG) \cite{tian2025chartgpt, patel2025starrystudioai, vaithilingam2024dynavis}.
    \item UI edits and modifications \cite{petridis2024insitu, liu2024usability}.
    \item Heuristic or accessibility analysis \cite{aljedaani2025accessibility, leiser2024hill, duan2024uifeedback}.
    \item Persona and content generation \cite{shaer2024integrating, bilgram2023aiinnovation}.
\end{itemize}

It seems that prompting has evolved into a design language, the ``lingua franca'' between designers and LLMs. We saw that LLMs are consistently deployed as semantic engines, processing abstract, natural language inputs to produce concrete outputs such as, HTML/CSS code \cite{wei2025aiui, kolthoff2024interlinking, patel2025starrystudioai}, prototypes and UI elements \cite{kolthoff2024interlinking, delatorre2024llmr}, design critiques and suggestions \cite{duan2024uifeedback, duan2024uicrit}, and user simulation and reasoning \cite{zhang2024designwatch, xiang2024simuser, benchaaben2024graphicalsyntax}. This prompt-based paradigm extends beyond traditional design tasks into novel UI behaviors, AI simulations of users, and interaction mapping, especially when combined with Chain-of-Thought (CoT) prompting, Few-shot learning, or multi-stage inference.

\subsubsection{Integration Across the Entire UI/UX Lifecycle}
LLMs are integrated into nearly every phase of the UI/UX design lifecycle, such as discovery and ideation, design generation, prototyping and simulation, evaluation and feedback, and iterative refinement. To better understand the practical impact of LLMs, we have mapped LLM integrations across the major stages of the UI/UX design lifecycle as illustrated in Table \ref{tab:llm_workflow_mapping}. This full-spectrum integration illustrates the maturation of LLMs as end-to-end design collaborators; they are no longer used only for ideation; they now support end-to-end UX workflows, from early research to post-prototype testing.

\begin{table*}[ht]
\centering
\caption{Mapping of LLM integration across UI/UX workflow stages}
\label{tab:llm_workflow_mapping}
\begin{tabular}{|p{3.5cm}|p{10.5cm}|}
\hline
\textbf{Design Stage} & \textbf{LLM Roles and Contributions} \\
\hline
\textbf{Research \& Discovery} & Summarizing user interviews, generating user personas, synthesizing survey responses, contextualizing user needs \cite {shaer2024integrating, wang2024aigc, takaffoli2024genaiux, bilgram2023aiinnovation}. \\
\hline
\textbf{Ideation} & Brainstorming design ideas, remixing initial concepts, exploring alternative interaction modalities, generating user stories \cite{shaer2024integrating, takaffoli2024genaiux, bilgram2023aiinnovation, wei2025aiui}. \\
\hline
\textbf{Design Generation} & LLMs are used to create mockups, widgets, layout code, and interaction logic from NL prompts \cite{patel2025starrystudioai, aljedaani2025accessibility, vaithilingam2024dynavis, masson2024directgpt, duan2024uifeedback}. \\
\hline
\textbf{Prototyping and Simulation} & Generating HTML/CSS code, creating UI mockups from natural language prompts, defining layout and behavior logic, enabling real-time editing \cite{patel2025starrystudioai, wei2025aiui, petridis2024insitu}. Tools like SimUser or LLMR simulate real users or interface behavior \cite{xiang2024simuser, delatorre2024llmr}. \\
\hline
\textbf{Evaluation and Feedback} & Performing heuristic critiques, identifying accessibility violations, simulating user interactions, providing structured feedback on usability, UI validation \cite{aljedaani2025accessibility, duan2024uifeedback, xiang2024simuser, leiser2024hill}. GPT-4, Gemini, and other models assess mockups using design heuristics or qualitative critique frameworks \cite{duan2024uifeedback, duan2024uicrit, duan2024uifeedback}. \\
\hline
\textbf{Iterative Refinement} &  Several systems (e.g., Farsight, DirectGPT) enable interactive loops, where users receive, refine, and re-prompt based on model output \cite{wang2024farsight, masson2024directgpt}. \\
\hline
\textbf{Reflection \& Ethics} & Detecting hallucinations and political bias, visualizing ethical risks, assessing potential harms, promoting responsible AI design \cite{wang2024farsight, leiser2024hill, benchaaben2024graphicalsyntax}. \\
\hline
\end{tabular}
\end{table*}

\subsubsection{Multimodality and Real-Time Interaction}
A growing number of studies involve multimodal LLMs (e.g., GPT-4V, Gemini Vision, PaLI), which process text + images/screenshots/video/audio to power more contextually rich and user-aware workflows. For instance, these multimodal LLMs are used to evaluate visual layouts \cite{wu2024uiclip, duan2024uicrit}, simulate user attention or cognitive flow \cite{zhang2024designwatch}, and annotate and respond to screenshots \cite{duan2024uicrit, wang2023conversationalui}. In immersive environments like VR and AR, LLMs are embedded in systems to support real-time behavior modification \cite{giunchi2024dreamcodevr}, augmented communication \cite{liu2023visualcaptions},  and context-aware storytelling and dialogue generation \cite{sasaki2024geofence}.

Vision-language processing is used in mobile prototyping \cite{petridis2024insitu}, design evaluation \cite{duan2024uicrit, wu2024uiclip}, and task planning from UI screenshots \cite{song2024visiontasker}. Hybrid models combine vision-based preprocessing (e.g., YOLO, OCR) with LLMs to enable semantic UI understanding \cite{song2024visiontasker}. These integrations show that the move toward context-aware AI agents that interact with full UI state, not just isolated text input, brings LLMs closer to cognitive co-designers. The fusion of LLMs with vision models and spatial interaction platforms marks the next frontier of dynamic UI/UX design. Table~\ref{tab:interaction_modalities} summarizes the modalities and associated studies. 

\begin{table*}[ht]
\centering
\caption{Interaction modalities of LLMs in UI/UX design workflows}
\label{tab:interaction_modalities}
\begin{tabular}{|p{4cm}|p{9cm}|}
\hline
\textbf{Modality} & \textbf{Description and Examples} \\
\hline
\textbf{Textual Prompting} & Natural language inputs guide LLM outputs for ideation \cite{shaer2024integrating, wang2024aigc}, code generation \cite{patel2025starrystudioai, wei2025aiui}, and heuristic analysis \cite{duan2024uifeedback}. \\
\hline
\textbf{Multimodal Inputs} & Prompts enriched with screenshots, user tasks, or video data for context-aware outputs \cite{zhang2024designwatch, wu2024uiclip, duan2024uicrit, song2024visiontasker}. \\
\hline
\textbf{Direct Manipulation Interfaces} & LLMs integrated into GUIs for interaction through drag-and-drop, buttons, or sliders.  \cite{masson2024directgpt, vaithilingam2024dynavis}. \\
\hline
\textbf{Voice-Based and Real-Time Systems} & Speech-to-text pipelines used in VR or conferencing platforms for live interface updates \cite{giunchi2024dreamcodevr, liu2023visualcaptions}. Context-aware storytelling and dialogue generation \cite{sasaki2024geofence}. \\
\hline
\end{tabular}
\end{table*}

\subsubsection{Modular and Iterative Workflows}
LLMs are increasingly embedded within modular workflows, where tasks are broken into interpretable subcomponents that the LLM processes in stages. These workflows are multistage and iterative \cite{tian2025chartgpt, wei2025aiui, kolthoff2024interlinking, benchaaben2024graphicalsyntax}, interactive and human-in-the-loop, where designers refine outputs via feedback \cite{zhou2024instructpipe, duan2024uifeedback, bilgram2023aiinnovation}, and visual + textual, combining language prompting with GUI manipulation \cite{vaithilingam2024dynavis, masson2024directgpt, delatorre2024llmr}.

Several tools enable in-situ design modification \cite{petridis2024insitu, vaithilingam2024dynavis}, interactive feedback visualization \cite{zhang2024designwatch, duan2024uicrit, duan2024uifeedback}, and back-and-forth refinement via dialog \cite{wei2025aiui, xiang2024simuser, takaffoli2024genaiux}. In this case, the LLM acts less as a static generator and more as a co-creative agent that supports exploration, iteration, and rapid prototyping.

\subsubsection{Human-AI Collaboration and Accessibility}
Another trend we observed is the use of LLMs in human-in-the-loop systems that promote responsible design, especially around accessibility, harm identification, and design ethics \cite{aljedaani2025accessibility, wang2024farsight, leiser2024hill}. LLMs assist designers in evaluating accessibility violations \cite{aljedaani2025accessibility}, simplify interfaces for non-technical users \cite{patel2025starrystudioai, petridis2024insitu, liu2024usability, takaffoli2024genaiux}, support inclusive design, like accessibility remediation \cite{aljedaani2025accessibility}, and simulate underrepresented personas (e.g., elderly users) in testing \cite{xiang2024simuser}.

Systems like Farsight augment design tools with widgets for surfacing harm-related risks \cite{wang2024farsight}. LLMs assess hallucination risks and political bias in generated UI outputs \cite{leiser2024hill}. There is also growing use of Wizard-of-Oz studies and user-in-the-loop mechanisms to evaluate AI output and refine interactions \cite{leiser2024hill, bilgram2023aiinnovation}. The trend here is emphasis is shifting toward democratizing design, enabling broader participation through natural language interfaces and inclusive evaluation.

\subsection{\textit{RQ2: What are the best practices identified in the literature for effectively integrating LLMs into UI/UX design processes?}}

The systematic review reveals a constellation of best practices that consistently emerge across the literature on integrating LLMs into UI/UX workflows. These practices cluster into six key themes: prompt engineering, human-in-the-loop iteration, tool integration, modularity, multimodal context-awareness, and trust and evaluation mechanisms.

\subsubsection{Prompt Engineering as a Design Tool}
A dominant trend across the reviewed studies is the critical role of prompt engineering in effectively integrating LLMs into design workflows. Prompting is not a one-off activity but an iterative, design-centric process. Many studies \cite{tian2025chartgpt, wei2025aiui, zamfirescu2023herding, wang2023conversationalui, bilgram2023aiinnovation} emphasize structured prompting strategies such as chain-of-thought reasoning, modular prompts aligned to interface elements, and few-shot learning using curated task exemplars. For instance, Duan et al. \cite{duan2024uicrit} introduce a two-stage prompting approach that separates critique generation from visual localization, enhancing precision and modularity. Similarly, Wei et al. \cite{wei2025aiui} and Zamfirescu-Pereira et al. \cite{zamfirescu2023herding} recommend crafting detailed, goal-oriented prompts enriched with user stories, example dialogues, or contextual variables to improve model relevance and coherence. Jiang et al. \cite{jiang2022promptmaker} note that teams benefit from collaborative prompt construction and the reuse of templates, supporting consistency and knowledge transfer within organizations.

\subsubsection{Human-in-the-Loop and Iterative Design Cycles}
Nearly all studies highlight the necessity of iterative human oversight. LLMs are most effective when used as \textit{co-creators} rather than autonomous agents. Across all the studies \cite{wei2025aiui,
tian2025chartgpt,
zhou2024instructpipe,
wang2024farsight,
takaffoli2024genaiux,
bilgram2023aiinnovation}, they recommend positioning the designer in the loop to edit, validate, and refine model outputs. For example,  Tian et al. \cite{tian2025chartgpt} outline editable pipelines where users can revise intermediate outputs, supporting correction and interpretability. Zhou et al. \cite{zhou2024instructpipe} similarly advocate prompt-based prototyping that allows manual tuning of LLM-generated content. Human-in-the-loop feedback loops, be they real-time \cite{giunchi2024dreamcodevr} or user study-driven \cite{patel2025starrystudioai}, consistently improve design alignment and usability. This practice not only enhances trust in AI outputs but also enables continuous learning for both the user and the model over time \cite{oruche2024holistic}.

\subsubsection{Integration with Existing Tools and Environments}
We identified that embedding LLMs within existing design environments facilitates smoother adoption. The studies \cite{patel2025starrystudioai, kolthoff2024interlinking, wang2024farsight, liu2023visualcaptions, petridis2023promptinfuser} describe how integration into familiar platforms such as Figma, Google Meet, or visual programming interfaces enhances accessibility and minimizes disruption. For instance, Patel et al.  \cite{patel2025starrystudioai} recommend seamless integration within widely adopted tools to ensure adoption by practitioners. Liu et al. \cite{liu2023visualcaptions}, focusing on Visual Captions, achieves real-time responsiveness by embedding LLM functionality into conversational interfaces with adjustable transparency, minimizing cognitive disruption. Kolthoff et al. \cite{kolthoff2024interlinking} highlights the importance of context-preserving UI transformations (e.g., structured abstraction layers) when embedding LLMs into prototyping workflows, which can reduce complexity and preserve interface fidelity.

\subsubsection{Modularity and Decomposition for Control and Scalability}
A recurring best practice involves breaking down UI/UX tasks into smaller, interpretable modules. Studies such as \cite{tian2025chartgpt, delatorre2024llmr, wang2023conversationalui} illustrate that LLMs perform more reliably when operating on abstracted, focused tasks. For example, Torre et al. \cite{delatorre2024llmr} advocate for modular LLM specialization, wherein individual agents (e.g., Planner GPT, Scene Analyzer) handle distinct subtasks such as planning, inspection, or debugging. Tian et al. \cite{tian2025chartgpt} propose decomposing tasks using least-to-most reasoning steps to improve control and performance. This approach allows designers to isolate model failures, reduce token bloat, and improve explainability. Modularity also supports scalability across workflows and user types, ensuring that systems can be adapted to a wide range of UI/UX applications.

\subsubsection{Multimodal Inputs and Context-Aware Interaction}
Several studies \cite{petridis2024insitu,
zhang2024designwatch,
xiang2024simuser,
wu2024uiclip,
benchaaben2024graphicalsyntax,
duan2024uifeedback} highlight the importance of grounding LLM outputs in multimodal or context-aware input formats. Zhang et al. \cite{zhang2024designwatch}, for instance, shows that combining textual task descriptions with UI screenshots results in more accurate and context-sensitive inferences. Wu et al. \cite{wu2024uiclip} go further by training LLMs using large-scale, visual design datasets (e.g., JitterWeb, BetterApp) to instill domain-specific principles such as contrast and alignment. Xiang et al. \cite{xiang2024simuser} leverages multimodal data, such as combining code with visual saliency models, to improve simulation fidelity and insight generation. These strategies ensure that the model maintains situational awareness of the user’s environment, device constraints, and interaction history, improving the robustness of LLM-generated recommendations.

\subsubsection{Evaluation, Explainability, and Trust-Building Mechanisms}
Trust in LLM-assisted workflows hinges on transparency, feedback mechanisms, and rigorous evaluation. Studies such as \cite{oruche2024holistic,
leiser2024hill,
duan2024uifeedback} emphasize the importance of explainability in both interface design and AI output interpretation. Leiser et al. \cite{leiser2024hill} introduce layered explanation mechanisms such as confidence scoring, bias indicators, and source attribution, all of which increase user trust and reduce the risk of hallucination. Oruche et al. \cite{oruche2024holistic} recommend XAI frameworks and standardized usability metrics (e.g., SUM, USE) to validate output quality. Duan et al. \cite{duan2024uifeedback}, meanwhile, includes features like dismissible suggestions and visual feedback for error localization. Collectively, these practices help make LLM behavior legible to users, aligning with core UX values of predictability, control, and feedback.

\begin{figure}
\centering 
\includegraphics[width=1\linewidth]{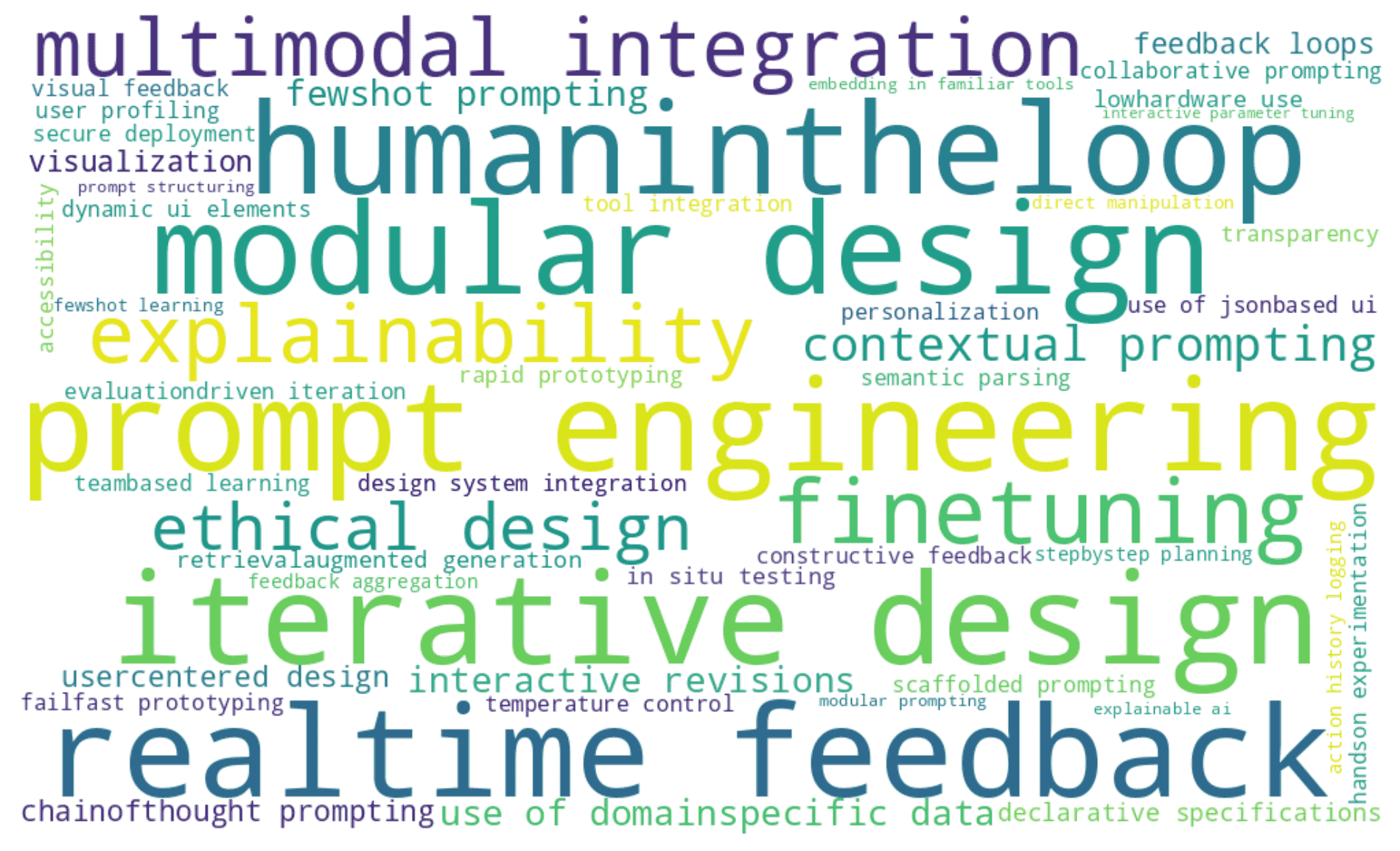}
\caption{Word cloud illustrating common best practices for integrating LLMs into UI/UX workflows from the reviewed studies.}
\label{fig:word-cloud}
\end{figure}

A qualitative overview of the most frequently mentioned best practices is illustrated in Figure \ref{fig:word-cloud}. The word cloud highlights dominant themes across studies, such as prompt engineering, human-in-the-loop design, tool integration, and modular workflows. These terms, appearing consistently across studies, reflect the centrality of human-centered, context-aware, and iterative strategies in LLM-augmented UI/UX processes. 

\begin{figure}
\centering 
\includegraphics[width=1\linewidth]{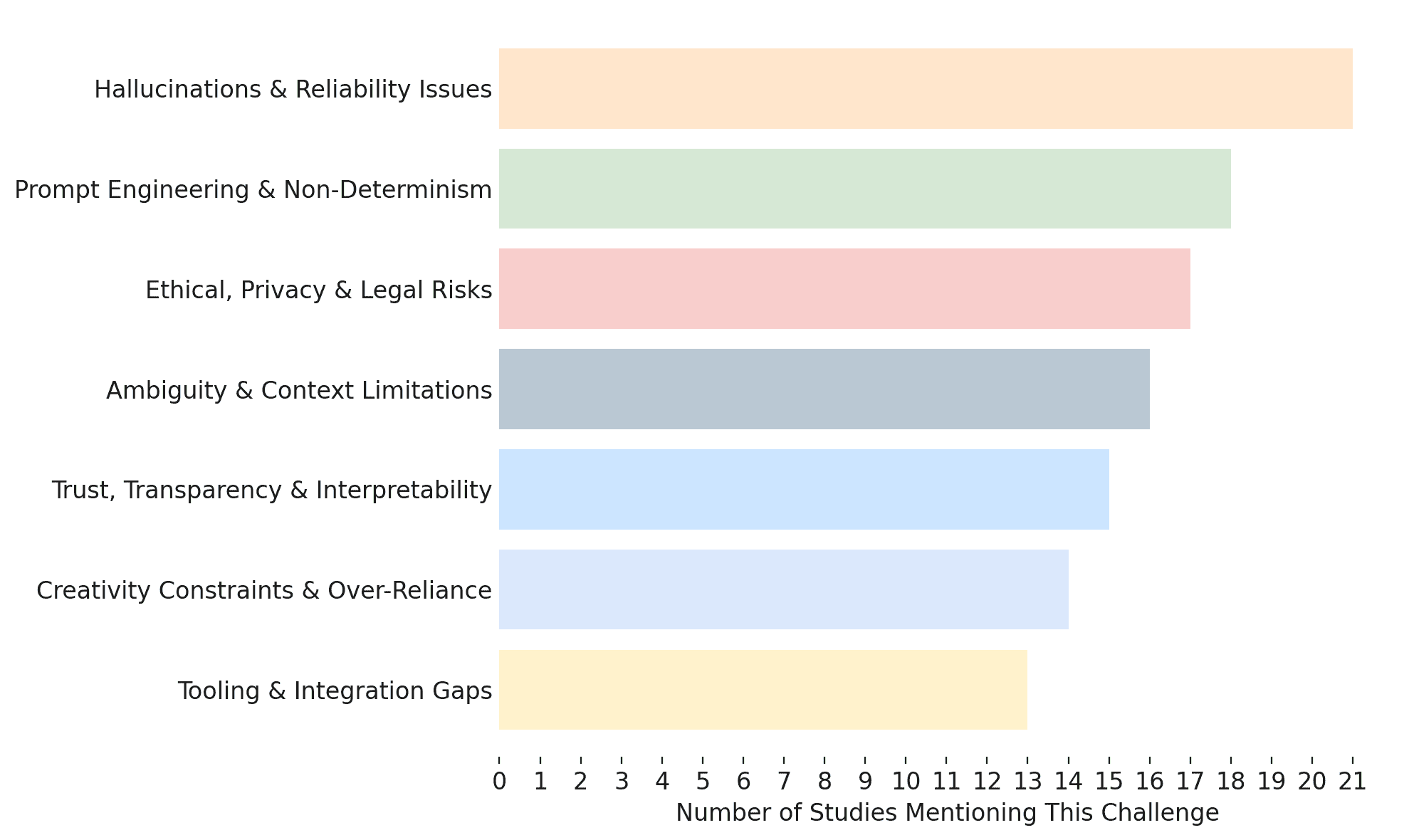}
\caption{Frequency of key challenges identified across studies on LLM integration in UI/UX workflows. Hallucinations, ethical concerns, and prompt-related issues emerged as the most commonly reported limitations.}
\label{fig:limitations}
\end{figure}

\subsection{RQ3: What are the primary limitations, challenges, and risks identified by researchers associated with integrating LLMs into UI/UX design workflows?}

A synthesis of the 38 reviewed studies reveals a diverse set of technical, human, and organizational challenges in integrating LLMs into UI/UX workflows. These limitations can be grouped into seven interrelated categories.

\subsubsection{Hallucinations, Inaccuracy, and Reliability Issues}
A recurring technical concern we observed is the tendency of LLMs to generate hallucinated content, such as fictional UI elements, flawed design critiques, or invalid code snippets, especially in underspecified or ambiguous scenarios~\cite{tian2025chartgpt, vaithilingam2024dynavis, leiser2024hill, duan2024uifeedback, duan2024uicrit, wang2023conversationalui}. This undermines trust and often requires manual verification to ensure design validity. Moreover, the inconsistent quality and accuracy of outputs reveal a fundamental fragility in current generative models.

\subsubsection{Prompt Engineering Challenges and Output Instability}
Studies highlight that LLMs are highly sensitive to prompt phrasing~\cite{aljedaani2025accessibility, shaer2024integrating, petridis2023promptinfuser, jiang2022promptmaker}. Creating effective prompts is a time-intensive process requiring experimentation, domain-specific knowledge, and iterative tuning. This problem is compounded by output non-determinism, where identical prompts can yield inconsistent results, limiting reproducibility and systematic iteration~\cite{tian2025chartgpt, zamfirescu2023herding}.

\subsubsection{Ambiguity, Context Loss, and Limited Multimodal Reasoning}
LLMs often struggle with incomplete or ambiguous prompts, particularly when required to interpret visual or spatial UI context~\cite{patel2025starrystudioai, masson2024directgpt, wang2023conversationalui, xiang2024simuser}. Token limitations and a lack of persistent memory constrain their ability to handle multi-screen workflows or interface dynamics~\cite{zhang2024designwatch, benchaaben2024graphicalsyntax}. Text-only models frequently underperform when interpreting layout, iconography, or hierarchical UI structures.

\subsubsection{Creativity Constraints and Over-Reliance Risks}
Several studies note that LLM outputs tend to converge on generic or conservative design patterns~\cite{shaer2024integrating, zhou2024instructpipe, wang2024aigc, benchaaben2024graphicalsyntax}, potentially limiting creative exploration. Designers may become anchored to AI-generated suggestions too early, reducing ideation breadth. There is also concern about over-reliance on LLMs, particularly among less experienced designers, leading to skill stagnation~\cite{wang2024aigc, takaffoli2024genaiux}.

\subsubsection{Trust, Transparency, and Interpretability}
LLMs often behave as black boxes. Designers frequently lack insight into how and why specific outputs are produced~\cite{oruche2024holistic, zhang2024designwatch, xiang2024simuser, kim2023cells}, making validation and debugging difficult. This opacity can create a false sense of precision, especially when presented within polished user interfaces~\cite{leiser2024hill, wang2024aigc}.

\subsubsection{Ethical, Privacy, and Legal Concerns}
The integration of LLMs introduces serious ethical and legal challenges. These include data privacy risks, unclear ownership of AI-generated content, and embedded biases in training data~\cite{wei2025aiui, oruche2024holistic, wang2024farsight, takaffoli2024genaiux, duan2024uifeedback}. A lack of governance frameworks, auditability, and inclusive datasets exacerbates concerns around accountability and fairness~\cite{singhal2024largeaction, lu2024aiisnotenough}.

\subsubsection{Tooling Gaps and Integration Limitations}
Several studies highlight a lack of seamless integration between LLMs and widely used design tools such as Figma or Sketch~\cite{kolthoff2024interlinking, takaffoli2024genaiux, bilgram2023aiinnovation}. Limitations in real-time collaboration, error recovery, and iteration tracking further hinder adoption. Hardware and deployment constraints also present practical challenges~\cite{wei2025aiui, singhal2024largeaction, zhang2024designwatch}.

The literature positions LLMs as powerful yet immature design collaborators. Their utility in accelerating ideation and generating interface elements is evident, but substantial limitations remain.

To address these challenges, future research and development should prioritize:

\begin{itemize}
    \item \textbf{Validation and Explainability}: Incorporating transparent model reasoning (e.g., confidence scores, rationale summaries) into LLM-driven tools to support debugging and trust.
    \item \textbf{Prompt Design Support}: Developing structured prompting templates, design grammars, or interactive co-pilots to reduce the cognitive burden of trial-and-error prompting.
    \item \textbf{Ethical Safeguards}: Embedding legal, ethical, and inclusivity checks early in the design workflow, supported by auditing and dataset governance.
    \item \textbf{Evaluation Standards}: Establishing benchmarks for creativity, usability, accessibility, and bias mitigation to assess LLM-generated designs in real-world contexts.
\end{itemize}

As shown in Figure~\ref{fig:limitations}, the most frequently cited challenges are hallucinations, ethical/privacy risks, and prompt-related instability. These limitations highlight the need for trustworthy, explainable, and ethically robust AI-assisted design systems.

\subsection{Reflections on Findings}
The studies reviewed in this work collectively point to an important shift in UI/UX design, where LLMs are no longer confined to back-end assistance but are becoming active participants in the creative process. Rather than being used solely for efficiency or automation, LLMs are increasingly embedded as co-creators, supporting ideation, critiquing interfaces, and even simulating users during testing. This transition from a hierarchical interaction model (designer commands, AI executes) to a dialogic one (designer and AI iterate together) is changing not only workflows but also the mindset of practitioners. It calls for a design practice that is more collaborative, open-ended, and comfortable with ambiguity.

Prompt engineering has emerged as a central, cross-cutting skill across studies. It is no longer a technical workaround; it’s becoming a creative discipline in itself. Designers are learning to treat prompts much like sketches or wireframes: iterative artifacts that can be tested, revised, and shared. Strategies like zero-shot and chain-of-thought prompting are being adapted into design routines, shaping how teams communicate intent to the model and interpret its responses.

Several systems reviewed demonstrate that vision-language models are enabling richer forms of interaction, where LLMs can understand not only what users say, but what they see, touch, or navigate through. This opens the door for context-aware design tools that can operate in real-world scenarios, including AR, VR, and mobile. These multimodal capacities reflect a growing expectation that design support should be responsive to both spatial and semantic context. At the same time, LLMs are lowering barriers to entry for non-technical users and supporting more inclusive outcomes. They are being used to simulate underserved user groups, flag accessibility issues, and help people with limited design or language skills contribute meaningfully to the process. However, this potential depends on intentional design. Who is empowered, and who is left out, still hinges on the assumptions baked into data, prompts, and interface choices.

While the speed and flexibility of LLMs offer clear benefits, particularly in early prototyping, there are risks to this acceleration. Several studies caution against rushing through design cycles without adequate reflection on ethics, transparency, and long-term impact. Responsible integration means building in checks: for hallucination detection, for harm anticipation, and for explainability.

Another recurring theme is fragmentation. Practices vary widely across teams, some using polished plugins within tools like Figma, others experimenting with bespoke systems or ad hoc prompting. This signals that the field is still in its early stages, lacking shared standards, frameworks, or even a common language for describing how LLMs should be evaluated or governed in design contexts.

Finally, the most effective applications tend to come from teams that blend technical expertise with deep domain knowledge. Custom models, tailored datasets, and domain-specific workflows aren’t just nice-to-haves; they’re essential for getting meaningful results. The next generation of design professionals will likely need to be hybrids: conversant in machine learning, fluent in design, and attentive to ethical complexity. Cross-functional collaboration isn’t just beneficial, it’s becoming foundational.

Taken together, these reflections suggest that language-driven, multimodal, and collaborative technologies are reshaping the future of UI/UX. The question is no longer whether to include LLMs in design workflows, but how to do so in ways that reflect our values, broaden participation, and maintain a strong sense of responsibility. As we move toward increasingly AI-augmented design processes, it will be crucial to strike a balance between speed and scrutiny, automation and agency, and capability and care.

\subsubsection{Implications for UI/UX Practitioners}

The integration of LLMs into UI/UX workflows is transforming not just how design tasks are carried out but also how designers conceptualize their role and process. Prompting is emerging as a core design skill, where crafting, refining, and reusing effective prompts becomes essential to guiding model behavior and achieving desired outcomes. Seamless tool integration is equally important, with LLMs proving most effective when embedded directly into platforms like Figma or Unity, thereby reducing friction and maintaining workflow continuity. Designers are encouraged to treat LLMs as collaborators rather than replacements, using AI-generated outputs as starting points for iteration, critique, and creative exploration. This shift calls for ethical awareness and explainability, as practitioners must be able to recognize biases, interpret uncertain outputs, and advocate for transparent AI tools. Importantly, LLMs also create opportunities for more inclusive and accessible design by simulating diverse perspectives and simplifying complexity for broader audiences. Given their probabilistic nature, workflows should be modular and adaptable, allowing for revision and in-the-loop human oversight. Beyond individual skills, organizational readiness is crucial; teams must invest in training, shared prompt resources, and cross-functional collaboration to leverage the benefits of LLMs fully. Collectively, these implications signal a broader evolution in design practice, one where designers act not just as creators but as facilitators of thoughtful, ethical, and collaborative human-AI design processes.
\section{Conclusion} \label{conclusion}

This systematic literature review examined how large language models (LLMs) are being integrated into UI/UX design workflows. Based on the analysis of 38 peer-reviewed studies, LLMs, particularly GPT-4 and its multimodal variants, are being used throughout the design lifecycle, including ideation, prototyping, evaluation, and refinement. Prompt-based interaction is the most common method of engagement, often supported by modular workflows, human-in-the-loop feedback, and integration with established design tools.

The review identifies several best practices, including structured prompting, iterative refinement, and context-aware multimodal inputs. These practices improve usability and support the effective deployment of LLMs, but they do not eliminate key challenges. Problems such as hallucinations, output instability, and limited explainability persist, which highlights the need for more transparent and reliable systems.

LLM integration in UI/UX remains fragmented, with no widely adopted standards or frameworks. Unlocking the full potential of these models will require interdisciplinary collaboration, stronger governance mechanisms, and a focus on responsible design. As these systems become more embedded in design processes, future work should focus not just on their capabilities but on ensuring that their use aligns with principles of transparency, inclusivity, and user-centered design.

\printcredits

\bibliographystyle{cas-model2-names}

\bibliography{Bibliography}

\section*{Appendix} \label{appendix}
Following is the list of studies from Figure \ref{fig:llm-integration-studies} along with their associated study entry number for reference:

\begin{enumerate}
    \item \cite{tian2025chartgpt}
    \item \cite{patel2025starrystudioai}
    \item \cite{wei2025aiui}
    \item \cite{aljedaani2025accessibility}
    \item \cite{petridis2024insitu}
    \item \cite{liu2024usability}
    \item \cite{kolthoff2024interlinking}
    \item \cite{oruche2024holistic}
    \item \cite{singhal2024largeaction}
    \item \cite{giunchi2024dreamcodevr}
    \item \cite{shaer2024integrating}
    \item \cite{zhou2024instructpipe}
    \item \cite{zhang2024designwatch}
    \item \cite{xiang2024simuser}
    \item \cite{wu2024uiclip}
    \item \cite{wang2024aigc}
    \item \cite{wang2024farsight}
    \item \cite{vaithilingam2024dynavis}
    \item \cite{takaffoli2024genaiux}
    \item \cite{song2024visiontasker}
    \item \cite{sasaki2024geofence}
    \item \cite{petridis2024promptinfuser}
    \item \cite{masson2024directgpt}
    \item \cite{lu2024aiisnotenough}
    \item \cite{leiser2024hill}
    \item \cite{duan2024uifeedback}
    \item \cite{duan2024uicrit}
    \item \cite{delatorre2024llmr}
    \item \cite{benchaaben2024graphicalsyntax}
    \item \cite{zamfirescu2023herding}
    \item \cite{wang2023conversationalui}
    \item \cite{petridis2023promptinfuser}
    \item \cite{liu2023visualcaptions}
    \item \cite{kim2023cells}
    \item \cite{duan2023uifeedback}
    \item \cite{du2023visualblocks}
    \item \cite{bilgram2023aiinnovation}
    \item \cite{jiang2022promptmaker}
\end{enumerate}

\end{document}